\documentclass[%
 reprint,
 amsmath,
 amssymb,
 aps,
]{revtex4-1}

\usepackage{siunitx}
\usepackage{graphicx}
\usepackage{braket}
\usepackage{verbatim}
\usepackage{dcolumn}
\usepackage{bm}
\usepackage{hyperref}

\begin{document}

\preprint{APS/123-QED}

\title{Engineering of the topological magnetic moment of electrons in bilayer graphene using strain and electrical bias}

\author{Christian Moulsdale}
\email{christian.moulsdale@postgrad.manchester.ac.uk}
\affiliation{School of Physics and Astronomy, University of Manchester, Manchester M13 9PL, UK}%
\affiliation{National Graphene Institute, University of Manchester, Manchester M13 9PL, UK}%

\author{Angelika Knothe}
\affiliation{School of Physics and Astronomy, University of Manchester, Manchester M13 9PL, UK}%
\affiliation{National Graphene Institute, University of Manchester, Manchester M13 9PL, UK}%

\author{Vladimir Fal'ko}
\affiliation{School of Physics and Astronomy, University of Manchester, Manchester M13 9PL, UK}%
\affiliation{National Graphene Institute, University of Manchester, Manchester M13 9PL, UK}%

\date{\today}

\begin{abstract}
Topological properties of electronic states in multivalley two-dimensional materials, such as mono- and bilayer graphene, or thin films of rhombohedral graphite, give rise to various unusual magneto-transport regimes. Here, we investigate the tunability of the topological magnetic moment (related to the Berry curvature) of electronic states in bilayer graphene using strain and vertical bias. We show how one can controllably vary the valley $g$-factor of the band-edge electrons, $g_v^*$, across the range $10 < |g_v^*| < 200$, and we discuss the manifestations of the topological magnetic moment in the anomalous contribution towards the Hall conductivity and in the Landau level spectrum.
\end{abstract}

    \maketitle

\begin{figure}[!htb]
    \centering
    \includegraphics{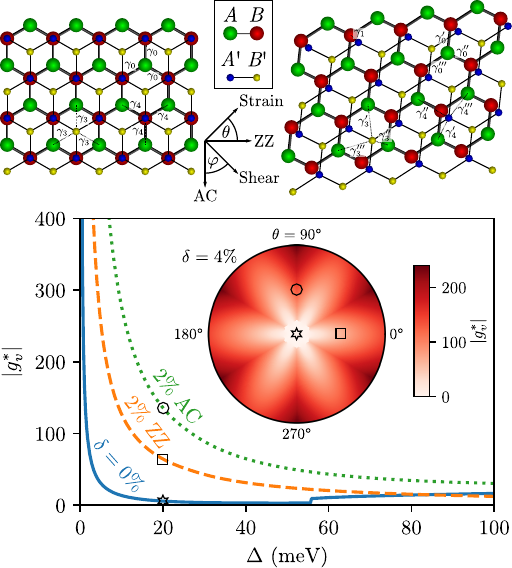}
    \caption{\textit{Top.} Unstrained (\textit{left}) and strained (\textit{right}) bilayer graphene (BLG) with the intra- and interlayer couplings $\gamma_{0,3,4}$ (modified by the strain) marked along the relevant hopping directions.  \textit{Bottom.} The magnitude of the valley $g$-factor at the conduction band edge of BLG, $|g_v^*|$, as a function of the interlayer asymmetry gap, $\Delta$, for uniaxial strains of magnitude ${\delta = 0\%}$ and 2\% applied along the zigzag (ZZ) and armchair (AC) directions. A jump in $|g_v^*|$ at $\Delta \sim \SI{55}{meV}$ for ${\delta = 0\%}$ is due to the disappearance of a central minivalley in the unstrained BLG spectrum upon the increase of the gap~\cite{PhysRevLett.96.086805}. \textit{Inset} shows $|g_v^*|$ against uniaxial strain (up to ${\delta = 4 \%}$) for various orientations of the strain tensor axes and ${\Delta = \SI{20}{meV}}$ (strain values used in the plot are marked by shapes). These images can also be used to characterize the effect of shear deformations described by Eq.~(\ref{eq:rho}) later in the text.}
    \label{fig:bilayer}
\end{figure}

\begin{figure*}[!htb]
    \centering
    \includegraphics{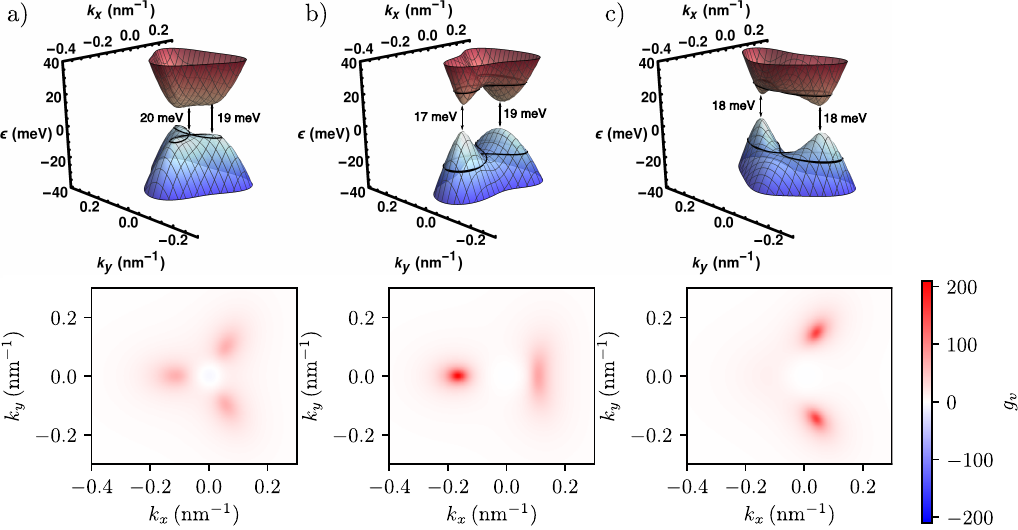}
    \caption{\textit{Top row}. The dispersion $\epsilon(\bm{k})$ of bilayer graphene in the $\bm{K}^-$ valley for ${\Delta = \SI{20}{meV}}$ and a) unstrained; b) 2\% uniaxial strain along the zigzag axis and c) 2\% uniaxial strain along the armchair axis (shear with parameters set by the relation in Eq.~(\ref{eq:rho})). Black lines indicate an energy cut at the van Hove singularity. \textit{Bottom row}. Contour plots of the corresponding valley $g$-factor $g_v$.}
    \label{fig:bands}
\end{figure*}

Strain in bilayer graphene (BLG), sketched in Fig.~\ref{fig:bilayer}, affects its low-energy electronic properties far greater than in its monolayer allotrope~\cite{PhysRevB.84.041404,PhysRevB.84.155410,MUCHAKRUCZYNSKI20111088,Wang2019,1909.13484}, generating qualitative changes in its low-energy spectrum close to the neutrality point. The earlier-discussed effects~\cite{PhysRevB.84.041404,PhysRevB.84.155410,MUCHAKRUCZYNSKI20111088} of unilateral strain and shear deformations in Bernal $(A'B)$ stacked bilayers include the  Lifshitz transition~\cite{lifshitz1960anomalies} for weakly n-doped and p-doped structures, accompanied by a redistribution (even a coalescence) of the Berry phase $\pm \pi$ singularities in the bilayer's electronic bands~\cite{PhysRevLett.96.086805,PhysRevB.84.041404}. These changes are caused by the interplay between the intralayer and skew $(AB')$ interlayer hopping parameters of electrons, modified by the deformations.

A transverse displacement field, induced by electrostatic gating of bilayers, is another factor that qualitatively changes their electronic properties. The displacement field generates an asymmetry between the layers, opening up a gap in the energy spectrum~\cite{PhysRevLett.96.086805,PhysRevB.74.161403} and smearing the Berry phase singularities into ``hot spots'' of Berry curvature, $\bm{\Omega}_\pm (\bm{p})$, located near the valley centers $\bm{K}^\pm$ (sign-inverted distributions are found in opposite valleys, $\bm{\Omega}_+ (\bm{p}) = \bm{\Omega}_- (- \bm{p})$). According to the fundamental properties of Bloch-Wannier functions~\cite{PhysRevB.53.7010,RevModPhys.82.1959}, a finite Berry curvature of the electronic bands is associated with a finite intrinsic angular momentum, therefore, a resulting magnetic moment of the plane-wave states of the electrons in the corresponding parts of the Brillouin zone (BZ) of the material~\cite{PhysRevB.53.7010,RevModPhys.82.1959,PhysRevB.98.155435}. The experimental signatures of such topological magnetic moments (TMM), with anomalously large effective $g$-factors $(g_v \sim 10 - 100)$, have recently been predicted~\cite{PhysRevB.98.155435} and, consequently, observed experimentally~\cite{PhysRevLett.121.257702,PhysRevLett.121.257703,PhysRevX.8.031023,1911.05968} in the studies of magnetotransport characteristics of electrostatically controlled wires in bilayer graphene. The Berry curvature and related intrinsic angular momentum are also associated with a ``Hall-like'' drift of electrons in a direction perpendicular to an external electric field, which causes topological valley currents~\cite{RevModPhys.82.1539} at ${B = 0}$ and an anomalous contribution toward the Hall conductivity of a 2D material subjected to an external magnetic field~\cite{1905.13094}.

In this paper, we study the interplay between strain and the interlayer asymmetry gap~\cite{PhysRevLett.96.086805} in BLG in determining topological properties of electronic states, such as the Berry curvature, recently analyzed in~\cite{PhysRevLett.123.036806}, and the topological magnetic moment, and their manifestations in the magnetotransport characteristics and Landau level spectra of bilayers. The outcome of this analysis is summarized in Fig.~\ref{fig:bilayer}, where we show how strain and shear increase the size of the effective valley $g$-factor for electrons and holes near the respective band edges of the gapped BLG, $g_v^*$, giving rise to its tunability by two orders of magnitude.

\begin{figure}[!htb]
    \centering
    \includegraphics{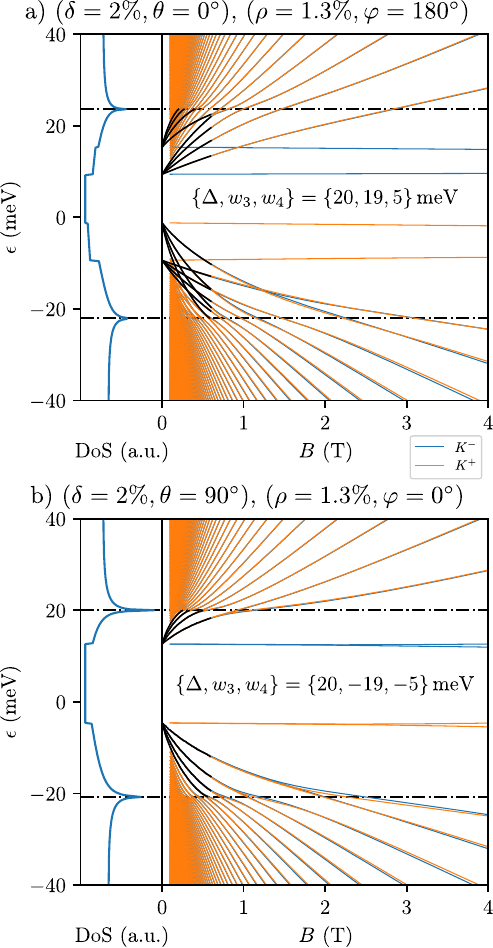}
    \caption{The zero magnetic field density of states (DoS) and numerically calculated Landau levels of bilayer graphene for ${B > \SI{0.1}{T}}$ in the $\bm{K}^+$ (orange) and $\bm{K}^-$ (blue) valleys of bilayer graphene with an interlayer asymmetry ${\Delta = \SI{20}{meV}}$ and a) 2\% uniaxial zigzag strain and b) 2\% uniaxial armchair strain (or related shear with parameters set by the relation in Eq.~(\ref{eq:rho})). The van Hove singularities are highlighted as black dot-dashed lines. A semiclassical approximation~\cite{LLc} for Landau levels near the band edges in Fig.~\ref{fig:bands}b) and c) (black lines) is used to extrapolate to ${B = 0}$. Note that the two-fold degeneracy of LLs at lower energies in b) is unique to the armchair direction of strain for which the spectrum in Fig.~\ref{fig:bands}c) features two degenerate minivalleys. This is lifted in the vicinity of the van Hove singularities. For an arbitrary orientation of the strain axes, the minivalleys are not degenerate even at ${B = 0}$.}
    \label{fig:LL}
\end{figure}

To describe electrons in the $\bm{K}^\xi$ $(\xi = \pm)$ valley of bilayers, we use the low-energy Hamiltonian written in the $(A, B', A', B)$ sublattice basis (marked in Fig.~\ref{fig:bilayer}),
\begin{align}
    \label{eq:H}
    &H_\xi =
    \begin{pmatrix}
        -\frac{1}{2} \Delta & v_3 \pi + w_3 & -v_4 \pi^\dag - w_4^* & v_0 \pi^\dag \\
        v_3 \pi^\dag + w_3^* & \frac{1}{2} \Delta & v_0 \pi & -v_4 \pi - w_4 \\
        -v_4 \pi - w_4 & v_0 \pi^\dag & \frac{1}{2} \Delta + \delta \epsilon & \gamma_1 \\
        v_0 \pi & -v_4 \pi^\dag - w_4^* & \gamma_1 & -\frac{1}{2} \Delta + \delta \epsilon
    \end{pmatrix}; \\
    \nonumber
    &w_{j=3,4} = \frac{3}{4} [ e^{-i2 \xi \theta} (\delta-\delta') (\eta_j-\eta_0) + 2 \sqrt{3} (-1) ^ j e^{i \xi \varphi} \rho \eta_j ] \gamma_j.
\end{align}
Here, $\pi = \xi p_x + i p_y$  and ${v_{0,3,4} = \sqrt{3} a\gamma_{0,3,4}/2\hbar}$ are determined by the intra- $(\gamma_0)$ and interlayer $(\gamma_{1,3,4})$ Slonczewski-Weiss hopping parameters~\cite{SW}, marked on the bilayer lattice in Fig.~\ref{fig:bilayer}, and $a$ is the lattice constant. For completeness, we take into account the dimer asymmetry $\delta \epsilon$ which, together with $\gamma_4$, breaks the particle-hole symmetry of the spectrum. The interlayer asymmetry, $\Delta = - e E_z d$, is induced by a transverse electric field $E_z$, where $d$ is the interlayer distance and $e < 0$ is the electron charge. The effect of strain is incorporated in Eq.~(\ref{eq:H}) in the form of gauge fields $w_{3,4}$, with the magnitude of $w_3$ partly enhanced by the Coulomb interaction~\cite{PhysRevB.84.041404,MUCHAKRUCZYNSKI20111088}. These come from the directional dependence of the $\gamma_{0,3,4}$ couplings~\cite{PhysRevB.84.041404,PhysRevB.84.155410,doi:10.1143/JPSJ.75.124701}, generated by strain and shear (relative shift of the layers). Note that the vertical $\gamma_1$ coupling is unaffected to first order in the strain amplitude. Here, $\theta$ is the angle between the zigzag crystallographic direction in graphene and the principal axis of the the strain tensor with components $\delta$ and $\delta' = -0.165 \delta$~\cite{doi:10.1021/acs.jpcc.8b04502}; shear deformations are described by $\rho = \delta r / a$, the interlayer lattice shift normalized by the lattice constant, where $\varphi$ is the angle between the shear direction and the armchair axis. Their effect is quantified using the Gr\"uneisen parameters ${\eta_j = \frac{r_{AB}}{\gamma_j} \frac{\partial \gamma_j}{\partial r_{AB}}}$, with the values ${\eta_0 \sim -3}$~\cite{PhysRevB.79.205433,doi:10.1021/nl101533x} and ${\eta_{3,4} \sim -1}$ taken from the literature~\cite{PhysRevB.85.125403}. Note that an uniaxial strain of magnitude $\delta$ at an angle $\theta$ is approximately equivalent to the shear deformation of magnitude $\rho$ and direction $\varphi$,
\begin{equation}
    \label{eq:rho}
    \rho = 0.336\delta \frac{\eta_0-\eta_3}{\eta_0}; \quad \varphi = 180^\circ - 2\theta.
\end{equation}
Note that for ${|w_4| < |w_3| \ll \gamma_1}$ the effects of $w_4$ on the spectral properties and kinetic parameters of electrons in a BLG is negligible.

Without any strain and for $\Delta = 0$, the BLG spectrum in the $\bm{K}^\pm$ valley~\cite{PhysRevB.77.113407} features a central Dirac cone with a Berry phase $\mp \pi$, surrounded by three Dirac points with Berry phase $\pm \pi$ (giving a total topological charge of $\pm 2 \pi$). Strain causes the displacement in the momentum plane and a coalescence of $\pm \pi$ singularities, whereas opening up a gap spreads them into hot spots of Berry curvature~\cite{RevModPhys.82.1959,PhysRevLett.123.196403,PhysRevLett.123.036806},
\begin{equation}
    \label{eq:Omega}
    \Omega_n^z = -2 \hbar^2 \, \mathrm{Im} \sum_{m \neq n} \frac{\braket{n|\partial_{p_x}H|m} \braket{m|\partial_{p_y}H|n}}{(\epsilon_n - \epsilon_m)^2}.
\end{equation}
In Fig.~\ref{fig:bands}, we illustrate the cumulative effect of the strain ($\delta = 2\%$, corresponding to ${|w_3| = \SI{19}{meV}}$ and ${|w_4| = \SI{5}{meV}}$) and a gap $(\Delta = \SI{20}{meV})$ on the BLG spectrum in the $\bm{K}^-$ valley, where the role of strain is to deform the three minivalleys at the BLG band edge~\cite{PhysRevLett.96.086805,PhysRevLett.113.116602} into two~\cite{PhysRevB.84.041404,PhysRevB.84.155410,PhysRevLett.123.196403}. Moreover, the particle-hole symmetry breaking caused by the hopping $\gamma_4$ and energy shift $\delta \epsilon$ of the dimer orbitals makes the bilayer band gap indirect for strain applied along the zigzag direction, as marked on Fig.~\ref{fig:bands}b). For strain applied along the armchair axis, the band edges remain degenerate in Fig.~\ref{fig:bands}c). In the bottom panels of Fig.~\ref{fig:bands}, we show the variation of the topological magnetic moment of the plane wave states of electrons across the Brillouin zone, computed using the relation derived in Ref.~~\cite{RevModPhys.82.1959},
\begin{equation}
    \label{eq:mu}
    \mu_n^z = -e \hbar \, \mathrm{Im} \sum_{m \neq n} \frac{\braket{n|\partial_{p_x}H|m} \braket{m|\partial_{p_y}H|n}}{\epsilon_n - \epsilon_m} \equiv \xi \mu_B g_v,
\end{equation}
where, for a band $n$, we sum across all three other bands $m \neq n$. This parameter reflects the valley splitting induced by simultaneous inversion (by $E_z$) and time-inversion (by $B$) symmetry breaking. We express $\mu^z$ in units of the Bohr magneton $\mu_B$, hence, present its values in terms of the valley $g$-factor, $g_v$. The dependence of $g_v$ at the conduction band edge, $g_v^*$, on the gap size, for BLG with 2\% strain applied along the zigzag and armchair axis of graphene is shown in Fig.~\ref{fig:bilayer}, together with a map describing the variation of $g_v^*$ with the change of the strain magnitude and orientation of its axes. Using a 2-band model~\cite{Novoselov2006,PhysRevB.84.041404}, which can be solved analytically, we estimate~\cite{2band} that for $|w_3| \gtrsim \Delta$, $g_v^* \sim -10^2 \xi |w_3| / \Delta$, in a good agreement with the numerical result.

\begin{figure}[!htb]
    \centering
    \includegraphics{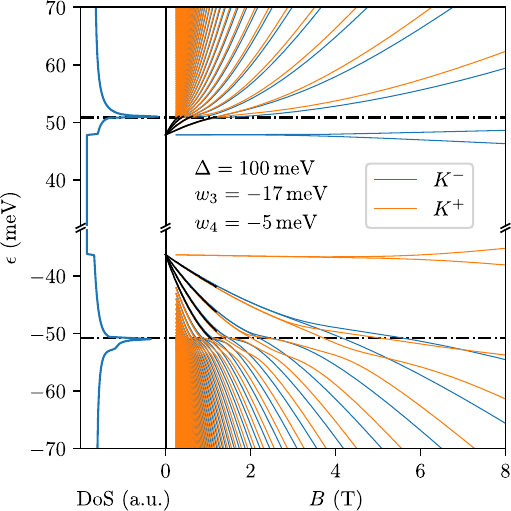}
    \caption{The density of states and numerically calculated Landau levels in both valleys for ${B > \SI{0.25}{T}}$ of bilayer graphene with an interlayer asymmetry ${\Delta = \SI{100}{meV}}$ and 2\% uniaxial armchair strain, with equivalent shear given by Eq.~(\ref{eq:rho}). The line convention is shared with Fig.~\ref{fig:LL}. Similarly to Fig.~\ref{fig:LL}b), Landau levels with energies below the saddle point are 2-fold degenerate in each valley.}
    \label{fig:LLK}
\end{figure}

\begin{figure*}[!htb]
    \centering
    \includegraphics{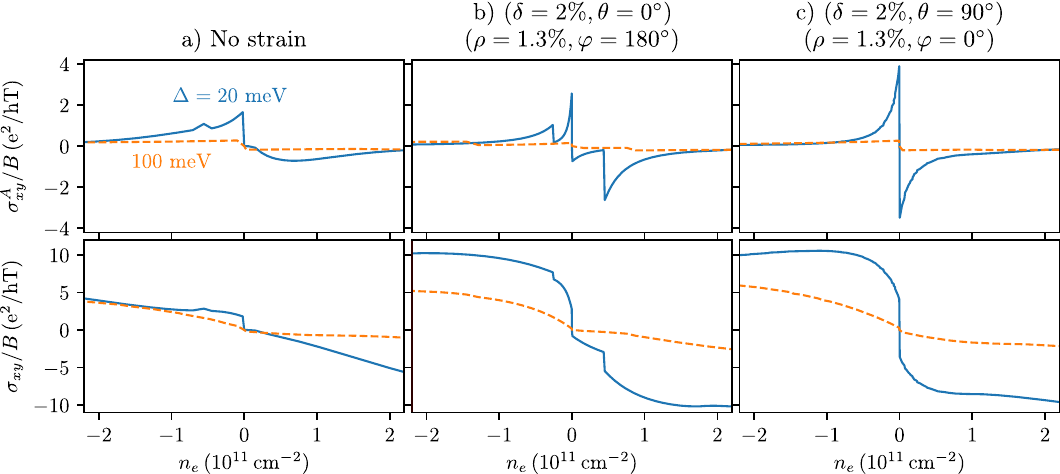}
    \caption{\textit{Top to bottom.} The anomalous Hall conductivity $\sigma_{xy}^A$ and total Hall conductivity $\sigma_{xy}$ of bilayer graphene for a small interlayer asymmetry ${\Delta = \SI{20}{meV}}$ (blue) and a large asymmetry $\SI{100}{meV}$ (orange) against carrier density $n_e$. \textit{Left to right.} No strain, 2\% uniaxial zigzag strain and 2\% uniaxial armchair strain.}
    \label{fig:sigma}
\end{figure*}

We also use the Hamiltonian in Eq.~(\ref{eq:H}) to compute~\cite{LL} the Landau level (LL) spectra in a simultaneously strained and vertically biased bilayer, in particular in the case of $\Delta \sim |w_3|$. In Fig.~\ref{fig:LL}, we show the LL spectra for a bilayer with ${\Delta = \SI{20}{meV}}$ and, due to 2\% uniaxial strain applied along the zigzag (ZZ) and armchair (AC) directions (or a ${\rho = 0.4 \%}$ shear anti-parallel and parallel to the AC axis according to Eq.~(\ref{eq:rho})), ${|w_3| = \SI{19}{meV}}$ and ${|w_4| = \SI{5}{meV}}$. Alongside this, we show the density of states (DoS) of the spectra at $B = 0$, marking van Hove singularities, and we extrapolate the LL spectra to the zero-field limit using semiclassical quantization~\cite{LLc} in the two minivalleys shown in Fig.~\ref{fig:bands}b) and c). While for an arbitrary orientation of the strain tensor axes these minivalleys are not degenerate, for strain applied along an armchair direction, the mirror symmetry of the crystal, retained despite the deformations, provides the degeneracy of the minivalleys. This also results in a double degeneracy of LLs at low energies which is lifted by magnetic breakdown that takes place at the saddle points in the spectra shown in Fig.~\ref{fig:bands}c). For the larger gap ${\Delta = \SI{100}{meV} \gg |w_3|}$, the effects of lattice deformations are diminished, so that the LL spectra in Fig.~\ref{fig:LLK} roughly coincide with what has been found earlier in gapped bilayers~\cite{PhysRevLett.113.116602}.

The formation of the topological magnetic moment may also manifest itself in the anomalous contribution towards the classical Hall effect in the bilayer. The latter is the result of a drift experienced by electrons in the bands with a finite $\bm{\Omega}$, in the direction perpendicular to the external electric field. Due to time-inversion symmetry, the resulting drift currents have the opposite signs in the opposite valleys $(\bm{K}^\pm)$, compensating each other at $B = 0$. However, a topological magnetic moment $\pm \mu^z$ leads to the splitting of BLG band edges between the $\bm{K}^\pm$ valleys, $\pm g_v \mu_B B_z$, and a valley contribution to the imbalance in the filling of the band edge state, leading to a finite anomalous Hall conductivity~\cite{RevModPhys.82.1539,1905.13094},
\begin{equation}
    \label{eq:sigmaA}
    \sigma_{xy}^A = -\frac{e^2 B}{\pi^2 \hbar^3} \sum_n \oint_{\epsilon_n (\bm{p}) = \epsilon_F} \frac{\Omega_n^z \mu_n^z}{|\bm{\nabla}_{\bm{p}} \epsilon_n|} dp.
\end{equation}
The former should be added to the classical Hall contribution~\cite{1905.13094},
\begin{equation}
    \label{eq:sigmaH}
    \sigma_{xy}^H = - \frac{e^3 \tau^2}{\pi^2 \hbar^2} \sum_{n,\gamma} \oint_{\epsilon_n (\bm{p}) = \epsilon_F} \frac{\partial_{p_x} \epsilon_n}{|\bm{\nabla}_{\bm{p}} \epsilon_n|} (\bm{\nabla}_{\bm{p}} \epsilon_n \times \bm{B})_\gamma \frac{dp}{m_{n, \gamma y}},
\end{equation}
where ${m_{n, \alpha \beta}^{-1} = \partial^2 \epsilon_n/\partial_{p_\alpha}\partial_{p_\beta}}$ and $\tau$ is the elastic scattering rate. The anomalous Hall conductivity originates from the TMM and is not suppressed by scattering, so its effect should be most pronounced in disordered bilayers. In Fig.~\ref{fig:sigma}, we show both $\sigma_{xy}^A$ and the total Hall conductivity ${\sigma_{xy} = \sigma_{xy}^H + \sigma_{xy}^A}$ against carrier density, $n_e$, for a strained and unstrained BLG with ${\tau \sim \SI{e-13}{s}}$~\cite{doi:10.1098/rsta.2007.2159}, and ${\Delta = \SI{20}{meV}}$ or $\SI{100}{meV}$.

Overall, the results presented above demonstrate that the topological characteristics of electron states in gapped bilayer graphene can be substantially enhanced by strain. This results in an anomalously large topological magnetic moment, leading to a valley splitting of band-edge states by magnetic field in a bias-gapped bilayer that leads to an anomalous correction to the Hall conductivity. The enhancement of topological effects may also be detected by measuring photocurrents induced by optically pumping bilayers using circularly polarized light similarly to that studied earlier in unstrained topological materials~\cite{1905.13094,Koppens}.

\textit{Acknowledgments}. We thank S. Slizovskiy, P. Makk, L. Wang, C. Stampfer, R. Danneau, F. Koppens, T. Ihn, and K. Ensslin for useful discussions. We acknowledge support from the EU Graphene Flagship Project; EPSRC Grants No. EP/L01548X/1, No. EP/S019367/1, No. EP/P026850/1, and No. EP/N010345/1; EC Project 2D-SIPC; and ERC Synergy Grant Hetero2D.

\bibliography{ref}

\begin{thebibliography}{36}%
\makeatletter
\providecommand \@ifxundefined [1]{%
 \@ifx{#1\undefined}
}%
\providecommand \@ifnum [1]{%
 \ifnum #1\expandafter \@firstoftwo
 \else \expandafter \@secondoftwo
 \fi
}%
\providecommand \@ifx [1]{%
 \ifx #1\expandafter \@firstoftwo
 \else \expandafter \@secondoftwo
 \fi
}%
\providecommand \natexlab [1]{#1}%
\providecommand \enquote  [1]{``#1''}%
\providecommand \bibnamefont  [1]{#1}%
\providecommand \bibfnamefont [1]{#1}%
\providecommand \citenamefont [1]{#1}%
\providecommand \href@noop [0]{\@secondoftwo}%
\providecommand \href [0]{\begingroup \@sanitize@url \@href}%
\providecommand \@href[1]{\@@startlink{#1}\@@href}%
\providecommand \@@href[1]{\endgroup#1\@@endlink}%
\providecommand \@sanitize@url [0]{\catcode `\\12\catcode `\$12\catcode
  `\&12\catcode `\#12\catcode `\^12\catcode `\_12\catcode `\%12\relax}%
\providecommand \@@startlink[1]{}%
\providecommand \@@endlink[0]{}%
\providecommand \url  [0]{\begingroup\@sanitize@url \@url }%
\providecommand \@url [1]{\endgroup\@href {#1}{\urlprefix }}%
\providecommand \urlprefix  [0]{URL }%
\providecommand \Eprint [0]{\href }%
\providecommand \doibase [0]{http://dx.doi.org/}%
\providecommand \selectlanguage [0]{\@gobble}%
\providecommand \bibinfo  [0]{\@secondoftwo}%
\providecommand \bibfield  [0]{\@secondoftwo}%
\providecommand \translation [1]{[#1]}%
\providecommand \BibitemOpen [0]{}%
\providecommand \bibitemStop [0]{}%
\providecommand \bibitemNoStop [0]{.\EOS\space}%
\providecommand \EOS [0]{\spacefactor3000\relax}%
\providecommand \BibitemShut  [1]{\csname bibitem#1\endcsname}%
\let\auto@bib@innerbib\@empty
\bibitem [{\citenamefont {McCann}\ and\ \citenamefont
  {Fal'ko}(2006)}]{PhysRevLett.96.086805}%
  \BibitemOpen
  \bibfield  {author} {\bibinfo {author} {\bibfnamefont {E.}~\bibnamefont
  {McCann}}\ and\ \bibinfo {author} {\bibfnamefont {V.~I.}\ \bibnamefont
  {Fal'ko}},\ }\href {\doibase 10.1103/PhysRevLett.96.086805} {\bibfield
  {journal} {\bibinfo  {journal} {Phys. Rev. Lett.}\ }\textbf {\bibinfo
  {volume} {96}},\ \bibinfo {pages} {086805} (\bibinfo {year}
  {2006})}\BibitemShut {NoStop}%
\bibitem [{\citenamefont {Mucha-Kruczy\ifmmode~\acute{n}\else \'{n}\fi{}ski}\
  \emph {et~al.}(2011)\citenamefont {Mucha-Kruczy\ifmmode~\acute{n}\else
  \'{n}\fi{}ski}, \citenamefont {Aleiner},\ and\ \citenamefont
  {Fal'ko}}]{PhysRevB.84.041404}%
  \BibitemOpen
  \bibfield  {author} {\bibinfo {author} {\bibfnamefont {M.}~\bibnamefont
  {Mucha-Kruczy\ifmmode~\acute{n}\else \'{n}\fi{}ski}}, \bibinfo {author}
  {\bibfnamefont {I.~L.}\ \bibnamefont {Aleiner}}, \ and\ \bibinfo {author}
  {\bibfnamefont {V.~I.}\ \bibnamefont {Fal'ko}},\ }\href {\doibase
  10.1103/PhysRevB.84.041404} {\bibfield  {journal} {\bibinfo  {journal} {Phys.
  Rev. B}\ }\textbf {\bibinfo {volume} {84}},\ \bibinfo {pages} {041404(R)}
  (\bibinfo {year} {2011})}\BibitemShut {NoStop}%
\bibitem [{\citenamefont {Son}\ \emph {et~al.}(2011)\citenamefont {Son},
  \citenamefont {Choi}, \citenamefont {Hong}, \citenamefont {Woo},\ and\
  \citenamefont {Jhi}}]{PhysRevB.84.155410}%
  \BibitemOpen
  \bibfield  {author} {\bibinfo {author} {\bibfnamefont {Y.-W.}\ \bibnamefont
  {Son}}, \bibinfo {author} {\bibfnamefont {S.-M.}\ \bibnamefont {Choi}},
  \bibinfo {author} {\bibfnamefont {Y.~P.}\ \bibnamefont {Hong}}, \bibinfo
  {author} {\bibfnamefont {S.}~\bibnamefont {Woo}}, \ and\ \bibinfo {author}
  {\bibfnamefont {S.-H.}\ \bibnamefont {Jhi}},\ }\href {\doibase
  10.1103/PhysRevB.84.155410} {\bibfield  {journal} {\bibinfo  {journal} {Phys.
  Rev. B}\ }\textbf {\bibinfo {volume} {84}},\ \bibinfo {pages} {155410}
  (\bibinfo {year} {2011})}\BibitemShut {NoStop}%
\bibitem [{\citenamefont {Mucha-Kruczyński}\ \emph {et~al.}(2011)\citenamefont
  {Mucha-Kruczyński}, \citenamefont {Aleiner},\ and\ \citenamefont
  {Fal’ko}}]{MUCHAKRUCZYNSKI20111088}%
  \BibitemOpen
  \bibfield  {author} {\bibinfo {author} {\bibfnamefont {M.}~\bibnamefont
  {Mucha-Kruczyński}}, \bibinfo {author} {\bibfnamefont {I.~L.}\ \bibnamefont
  {Aleiner}}, \ and\ \bibinfo {author} {\bibfnamefont {V.~I.}\ \bibnamefont
  {Fal’ko}},\ }\href {\doibase https://doi.org/10.1016/j.ssc.2011.05.019}
  {\bibfield  {journal} {\bibinfo  {journal} {Solid State Communications}\
  }\textbf {\bibinfo {volume} {151}},\ \bibinfo {pages} {1088 } (\bibinfo
  {year} {2011})}\BibitemShut {NoStop}%
\bibitem [{\citenamefont {Wang}\ \emph
  {et~al.}(2019{\natexlab{a}})\citenamefont {Wang}, \citenamefont {Zihlmann},
  \citenamefont {Baumgartner}, \citenamefont {Overbeck}, \citenamefont
  {Watanabe}, \citenamefont {Taniguchi}, \citenamefont {Makk},\ and\
  \citenamefont {Sch{\"o}nenberger}}]{Wang2019}%
  \BibitemOpen
  \bibfield  {author} {\bibinfo {author} {\bibfnamefont {L.}~\bibnamefont
  {Wang}}, \bibinfo {author} {\bibfnamefont {S.}~\bibnamefont {Zihlmann}},
  \bibinfo {author} {\bibfnamefont {A.}~\bibnamefont {Baumgartner}}, \bibinfo
  {author} {\bibfnamefont {J.}~\bibnamefont {Overbeck}}, \bibinfo {author}
  {\bibfnamefont {K.}~\bibnamefont {Watanabe}}, \bibinfo {author}
  {\bibfnamefont {T.}~\bibnamefont {Taniguchi}}, \bibinfo {author}
  {\bibfnamefont {P.}~\bibnamefont {Makk}}, \ and\ \bibinfo {author}
  {\bibfnamefont {C.}~\bibnamefont {Sch{\"o}nenberger}},\ }\href {\doibase
  10.1021/acs.nanolett.9b01491} {\bibfield  {journal} {\bibinfo  {journal}
  {Nano Letters}\ }\textbf {\bibinfo {volume} {19}},\ \bibinfo {pages} {4097}
  (\bibinfo {year} {2019}{\natexlab{a}})}\BibitemShut {NoStop}%
\bibitem [{\citenamefont {Wang}\ \emph
  {et~al.}(2019{\natexlab{b}})\citenamefont {Wang}, \citenamefont {Makk},
  \citenamefont {Zihlmann}, \citenamefont {Baumgartner}, \citenamefont
  {Indolese}, \citenamefont {Watanabe}, \citenamefont {Taniguchi},\ and\
  \citenamefont {Schönenberger}}]{1909.13484}%
  \BibitemOpen
  \bibfield  {author} {\bibinfo {author} {\bibfnamefont {L.}~\bibnamefont
  {Wang}}, \bibinfo {author} {\bibfnamefont {P.}~\bibnamefont {Makk}}, \bibinfo
  {author} {\bibfnamefont {S.}~\bibnamefont {Zihlmann}}, \bibinfo {author}
  {\bibfnamefont {A.}~\bibnamefont {Baumgartner}}, \bibinfo {author}
  {\bibfnamefont {D.~I.}\ \bibnamefont {Indolese}}, \bibinfo {author}
  {\bibfnamefont {K.}~\bibnamefont {Watanabe}}, \bibinfo {author}
  {\bibfnamefont {T.}~\bibnamefont {Taniguchi}}, \ and\ \bibinfo {author}
  {\bibfnamefont {C.}~\bibnamefont {Schönenberger}},\ }\href@noop {} {}
  (\bibinfo {year} {2019}{\natexlab{b}}),\ \Eprint
  {http://arxiv.org/abs/arXiv:1909.13484} {arXiv:1909.13484} \BibitemShut
  {NoStop}%
\bibitem [{\citenamefont {Lifshitz}\ \emph {et~al.}(1960)\citenamefont
  {Lifshitz} \emph {et~al.}}]{lifshitz1960anomalies}%
  \BibitemOpen
  \bibfield  {author} {\bibinfo {author} {\bibfnamefont {I.}~\bibnamefont
  {Lifshitz}} \emph {et~al.},\ }\href@noop {} {\bibfield  {journal} {\bibinfo
  {journal} {Sov. Phys. JETP}\ }\textbf {\bibinfo {volume} {11}},\ \bibinfo
  {pages} {1130} (\bibinfo {year} {1960})}\BibitemShut {NoStop}%
\bibitem [{\citenamefont {McCann}(2006)}]{PhysRevB.74.161403}%
  \BibitemOpen
  \bibfield  {author} {\bibinfo {author} {\bibfnamefont {E.}~\bibnamefont
  {McCann}},\ }\href {\doibase 10.1103/PhysRevB.74.161403} {\bibfield
  {journal} {\bibinfo  {journal} {Phys. Rev. B}\ }\textbf {\bibinfo {volume}
  {74}},\ \bibinfo {pages} {161403(R)} (\bibinfo {year} {2006})}\BibitemShut
  {NoStop}%
\bibitem [{\citenamefont {Chang}\ and\ \citenamefont
  {Niu}(1996)}]{PhysRevB.53.7010}%
  \BibitemOpen
  \bibfield  {author} {\bibinfo {author} {\bibfnamefont {M.-C.}\ \bibnamefont
  {Chang}}\ and\ \bibinfo {author} {\bibfnamefont {Q.}~\bibnamefont {Niu}},\
  }\href {\doibase 10.1103/PhysRevB.53.7010} {\bibfield  {journal} {\bibinfo
  {journal} {Phys. Rev. B}\ }\textbf {\bibinfo {volume} {53}},\ \bibinfo
  {pages} {7010} (\bibinfo {year} {1996})}\BibitemShut {NoStop}%
\bibitem [{\citenamefont {Xiao}\ \emph {et~al.}(2010)\citenamefont {Xiao},
  \citenamefont {Chang},\ and\ \citenamefont {Niu}}]{RevModPhys.82.1959}%
  \BibitemOpen
  \bibfield  {author} {\bibinfo {author} {\bibfnamefont {D.}~\bibnamefont
  {Xiao}}, \bibinfo {author} {\bibfnamefont {M.-C.}\ \bibnamefont {Chang}}, \
  and\ \bibinfo {author} {\bibfnamefont {Q.}~\bibnamefont {Niu}},\ }\href
  {\doibase 10.1103/RevModPhys.82.1959} {\bibfield  {journal} {\bibinfo
  {journal} {Rev. Mod. Phys.}\ }\textbf {\bibinfo {volume} {82}},\ \bibinfo
  {pages} {1959} (\bibinfo {year} {2010})}\BibitemShut {NoStop}%
\bibitem [{\citenamefont {Knothe}\ and\ \citenamefont
  {Fal'ko}(2018)}]{PhysRevB.98.155435}%
  \BibitemOpen
  \bibfield  {author} {\bibinfo {author} {\bibfnamefont {A.}~\bibnamefont
  {Knothe}}\ and\ \bibinfo {author} {\bibfnamefont {V.}~\bibnamefont
  {Fal'ko}},\ }\href {\doibase 10.1103/PhysRevB.98.155435} {\bibfield
  {journal} {\bibinfo  {journal} {Phys. Rev. B}\ }\textbf {\bibinfo {volume}
  {98}},\ \bibinfo {pages} {155435} (\bibinfo {year} {2018})}\BibitemShut
  {NoStop}%
\bibitem [{\citenamefont {Overweg}\ \emph {et~al.}(2018)\citenamefont
  {Overweg}, \citenamefont {Knothe}, \citenamefont {Fabian}, \citenamefont
  {Linhart}, \citenamefont {Rickhaus}, \citenamefont {Wernli}, \citenamefont
  {Watanabe}, \citenamefont {Taniguchi}, \citenamefont {S\'anchez},
  \citenamefont {Burgd\"orfer}, \citenamefont {Libisch}, \citenamefont
  {Fal'ko}, \citenamefont {Ensslin},\ and\ \citenamefont
  {Ihn}}]{PhysRevLett.121.257702}%
  \BibitemOpen
  \bibfield  {author} {\bibinfo {author} {\bibfnamefont {H.}~\bibnamefont
  {Overweg}}, \bibinfo {author} {\bibfnamefont {A.}~\bibnamefont {Knothe}},
  \bibinfo {author} {\bibfnamefont {T.}~\bibnamefont {Fabian}}, \bibinfo
  {author} {\bibfnamefont {L.}~\bibnamefont {Linhart}}, \bibinfo {author}
  {\bibfnamefont {P.}~\bibnamefont {Rickhaus}}, \bibinfo {author}
  {\bibfnamefont {L.}~\bibnamefont {Wernli}}, \bibinfo {author} {\bibfnamefont
  {K.}~\bibnamefont {Watanabe}}, \bibinfo {author} {\bibfnamefont
  {T.}~\bibnamefont {Taniguchi}}, \bibinfo {author} {\bibfnamefont
  {D.}~\bibnamefont {S\'anchez}}, \bibinfo {author} {\bibfnamefont
  {J.}~\bibnamefont {Burgd\"orfer}}, \bibinfo {author} {\bibfnamefont
  {F.}~\bibnamefont {Libisch}}, \bibinfo {author} {\bibfnamefont {V.~I.}\
  \bibnamefont {Fal'ko}}, \bibinfo {author} {\bibfnamefont {K.}~\bibnamefont
  {Ensslin}}, \ and\ \bibinfo {author} {\bibfnamefont {T.}~\bibnamefont
  {Ihn}},\ }\href {\doibase 10.1103/PhysRevLett.121.257702} {\bibfield
  {journal} {\bibinfo  {journal} {Phys. Rev. Lett.}\ }\textbf {\bibinfo
  {volume} {121}},\ \bibinfo {pages} {257702} (\bibinfo {year}
  {2018})}\BibitemShut {NoStop}%
\bibitem [{\citenamefont {Kraft}\ \emph {et~al.}(2018)\citenamefont {Kraft},
  \citenamefont {Krainov}, \citenamefont {Gall}, \citenamefont {Dmitriev},
  \citenamefont {Krupke}, \citenamefont {Gornyi},\ and\ \citenamefont
  {Danneau}}]{PhysRevLett.121.257703}%
  \BibitemOpen
  \bibfield  {author} {\bibinfo {author} {\bibfnamefont {R.}~\bibnamefont
  {Kraft}}, \bibinfo {author} {\bibfnamefont {I.~V.}\ \bibnamefont {Krainov}},
  \bibinfo {author} {\bibfnamefont {V.}~\bibnamefont {Gall}}, \bibinfo {author}
  {\bibfnamefont {A.~P.}\ \bibnamefont {Dmitriev}}, \bibinfo {author}
  {\bibfnamefont {R.}~\bibnamefont {Krupke}}, \bibinfo {author} {\bibfnamefont
  {I.~V.}\ \bibnamefont {Gornyi}}, \ and\ \bibinfo {author} {\bibfnamefont
  {R.}~\bibnamefont {Danneau}},\ }\href {\doibase
  10.1103/PhysRevLett.121.257703} {\bibfield  {journal} {\bibinfo  {journal}
  {Phys. Rev. Lett.}\ }\textbf {\bibinfo {volume} {121}},\ \bibinfo {pages}
  {257703} (\bibinfo {year} {2018})}\BibitemShut {NoStop}%
\bibitem [{\citenamefont {Eich}\ \emph {et~al.}(2018)\citenamefont {Eich},
  \citenamefont {Herman}, \citenamefont {Pisoni}, \citenamefont {Overweg},
  \citenamefont {Kurzmann}, \citenamefont {Lee}, \citenamefont {Rickhaus},
  \citenamefont {Watanabe}, \citenamefont {Taniguchi}, \citenamefont {Sigrist},
  \citenamefont {Ihn},\ and\ \citenamefont {Ensslin}}]{PhysRevX.8.031023}%
  \BibitemOpen
  \bibfield  {author} {\bibinfo {author} {\bibfnamefont {M.}~\bibnamefont
  {Eich}}, \bibinfo {author} {\bibfnamefont {F.}~\bibnamefont {Herman}},
  \bibinfo {author} {\bibfnamefont {R.}~\bibnamefont {Pisoni}}, \bibinfo
  {author} {\bibfnamefont {H.}~\bibnamefont {Overweg}}, \bibinfo {author}
  {\bibfnamefont {A.}~\bibnamefont {Kurzmann}}, \bibinfo {author}
  {\bibfnamefont {Y.}~\bibnamefont {Lee}}, \bibinfo {author} {\bibfnamefont
  {P.}~\bibnamefont {Rickhaus}}, \bibinfo {author} {\bibfnamefont
  {K.}~\bibnamefont {Watanabe}}, \bibinfo {author} {\bibfnamefont
  {T.}~\bibnamefont {Taniguchi}}, \bibinfo {author} {\bibfnamefont
  {M.}~\bibnamefont {Sigrist}}, \bibinfo {author} {\bibfnamefont
  {T.}~\bibnamefont {Ihn}}, \ and\ \bibinfo {author} {\bibfnamefont
  {K.}~\bibnamefont {Ensslin}},\ }\href {\doibase 10.1103/PhysRevX.8.031023}
  {\bibfield  {journal} {\bibinfo  {journal} {Phys. Rev. X}\ }\textbf {\bibinfo
  {volume} {8}},\ \bibinfo {pages} {031023} (\bibinfo {year}
  {2018})}\BibitemShut {NoStop}%
\bibitem [{\citenamefont {Lee}\ \emph {et~al.}(2019)\citenamefont {Lee},
  \citenamefont {Knothe}, \citenamefont {Rickhaus}, \citenamefont {Overweg},
  \citenamefont {Eich}, \citenamefont {Kurzmann}, \citenamefont {Taniguchi},
  \citenamefont {Wantanabe}, \citenamefont {Fal'ko}, \citenamefont {Ihn},\ and\
  \citenamefont {Ensslin}}]{1911.05968}%
  \BibitemOpen
  \bibfield  {author} {\bibinfo {author} {\bibfnamefont {Y.}~\bibnamefont
  {Lee}}, \bibinfo {author} {\bibfnamefont {A.}~\bibnamefont {Knothe}},
  \bibinfo {author} {\bibfnamefont {P.}~\bibnamefont {Rickhaus}}, \bibinfo
  {author} {\bibfnamefont {H.}~\bibnamefont {Overweg}}, \bibinfo {author}
  {\bibfnamefont {M.}~\bibnamefont {Eich}}, \bibinfo {author} {\bibfnamefont
  {A.}~\bibnamefont {Kurzmann}}, \bibinfo {author} {\bibfnamefont
  {T.}~\bibnamefont {Taniguchi}}, \bibinfo {author} {\bibfnamefont
  {K.}~\bibnamefont {Wantanabe}}, \bibinfo {author} {\bibfnamefont
  {V.}~\bibnamefont {Fal'ko}}, \bibinfo {author} {\bibfnamefont
  {T.}~\bibnamefont {Ihn}}, \ and\ \bibinfo {author} {\bibfnamefont
  {K.}~\bibnamefont {Ensslin}},\ }\href@noop {} {} (\bibinfo {year} {2019}),\
  \Eprint {http://arxiv.org/abs/arXiv:1911.05968} {arXiv:1911.05968}
  \BibitemShut {NoStop}%
\bibitem [{\citenamefont {Nagaosa}\ \emph {et~al.}(2010)\citenamefont
  {Nagaosa}, \citenamefont {Sinova}, \citenamefont {Onoda}, \citenamefont
  {MacDonald},\ and\ \citenamefont {Ong}}]{RevModPhys.82.1539}%
  \BibitemOpen
  \bibfield  {author} {\bibinfo {author} {\bibfnamefont {N.}~\bibnamefont
  {Nagaosa}}, \bibinfo {author} {\bibfnamefont {J.}~\bibnamefont {Sinova}},
  \bibinfo {author} {\bibfnamefont {S.}~\bibnamefont {Onoda}}, \bibinfo
  {author} {\bibfnamefont {A.~H.}\ \bibnamefont {MacDonald}}, \ and\ \bibinfo
  {author} {\bibfnamefont {N.~P.}\ \bibnamefont {Ong}},\ }\href {\doibase
  10.1103/RevModPhys.82.1539} {\bibfield  {journal} {\bibinfo  {journal} {Rev.
  Mod. Phys.}\ }\textbf {\bibinfo {volume} {82}},\ \bibinfo {pages} {1539}
  (\bibinfo {year} {2010})}\BibitemShut {NoStop}%
\bibitem [{\citenamefont {Slizovskiy}\ \emph {et~al.}(2019)\citenamefont
  {Slizovskiy}, \citenamefont {McCann}, \citenamefont {Koshino},\ and\
  \citenamefont {Fal'ko}}]{1905.13094}%
  \BibitemOpen
  \bibfield  {author} {\bibinfo {author} {\bibfnamefont {S.}~\bibnamefont
  {Slizovskiy}}, \bibinfo {author} {\bibfnamefont {E.}~\bibnamefont {McCann}},
  \bibinfo {author} {\bibfnamefont {M.}~\bibnamefont {Koshino}}, \ and\
  \bibinfo {author} {\bibfnamefont {V.~I.}\ \bibnamefont {Fal'ko}},\
  }\href@noop {} {} (\bibinfo {year} {2019}),\ \Eprint
  {http://arxiv.org/abs/arXiv:1905.13094} {arXiv:1905.13094} \BibitemShut
  {NoStop}%
\bibitem [{\citenamefont {Son}\ \emph {et~al.}(2019)\citenamefont {Son},
  \citenamefont {Kim}, \citenamefont {Ahn}, \citenamefont {Lee},\ and\
  \citenamefont {Lee}}]{PhysRevLett.123.036806}%
  \BibitemOpen
  \bibfield  {author} {\bibinfo {author} {\bibfnamefont {J.}~\bibnamefont
  {Son}}, \bibinfo {author} {\bibfnamefont {K.-H.}\ \bibnamefont {Kim}},
  \bibinfo {author} {\bibfnamefont {Y.~H.}\ \bibnamefont {Ahn}}, \bibinfo
  {author} {\bibfnamefont {H.-W.}\ \bibnamefont {Lee}}, \ and\ \bibinfo
  {author} {\bibfnamefont {J.}~\bibnamefont {Lee}},\ }\href {\doibase
  10.1103/PhysRevLett.123.036806} {\bibfield  {journal} {\bibinfo  {journal}
  {Phys. Rev. Lett.}\ }\textbf {\bibinfo {volume} {123}},\ \bibinfo {pages}
  {036806} (\bibinfo {year} {2019})}\BibitemShut {NoStop}%
\bibitem [{LLc()}]{LLc}%
  \BibitemOpen
  \href@noop {} {}\bibinfo {note} {The semiclassical quantization of a
  cyclotron orbit (isoenergy line of the eigenvalues of Eq.~(\ref{eq:H})) of
  energy $\epsilon$ is ${S(\epsilon) \ell_B^2 = \pi (n + 1) -
  \Gamma(\mathcal{\epsilon})}$, where $S(\epsilon)$ is its area in
  $\bm{k}$-space, $\ell_B = \sqrt{\hbar / |e|B}$ is the magnetic length,
  $\Gamma(\mathcal{\epsilon})$ is the Berry phase accumulated along the contour
  (0 for $\Delta > 0$) and $n$ is the index. For an explanation of this
  expression, see appendix A of \cite{Fuchs2010}.}\BibitemShut {Stop}%
\bibitem [{SW()}]{SW}%
  \BibitemOpen
  \href@noop {} {}\bibinfo {note} {$\gamma_0 = \braket{A|\mathcal{H}|B} =
  \braket{A'|\mathcal{H}|B'} = \SI{3.161}{eV}$, $\gamma_1 =
  \braket{A'|\mathcal{H}|B} = \SI{0.381}{eV}$, $\gamma_3 =
  \braket{A|\mathcal{H}|B'} = \SI{0.38}{eV}$, $\gamma_4 =
  \braket{A|\mathcal{H}|A'} = \braket{B|\mathcal{H}|B'} = \SI{0.14}{eV}$ and
  $\delta \epsilon = \braket{B|\mathcal{H}|B} - \braket{A|\mathcal{H}|A} =
  \braket{A'|\mathcal{H}|A'} - \braket{B'|\mathcal{H}|B'} = \SI{0.022}{eV}$.
  Here we use parameters proposed in
  Ref.~\cite{PhysRevB.80.165406}.}\BibitemShut {Stop}%
\bibitem [{\citenamefont {Ando}(2006)}]{doi:10.1143/JPSJ.75.124701}%
  \BibitemOpen
  \bibfield  {author} {\bibinfo {author} {\bibfnamefont {T.}~\bibnamefont
  {Ando}},\ }\href {\doibase 10.1143/JPSJ.75.124701} {\bibfield  {journal}
  {\bibinfo  {journal} {Journal of the Physical Society of Japan}\ }\textbf
  {\bibinfo {volume} {75}},\ \bibinfo {pages} {124701} (\bibinfo {year}
  {2006})}\BibitemShut {NoStop}%
\bibitem [{\citenamefont {Botello-M\'endez}\ \emph {et~al.}(2018)\citenamefont
  {Botello-M\'endez}, \citenamefont {Obeso-Jureidini},\ and\ \citenamefont
  {Naumis}}]{doi:10.1021/acs.jpcc.8b04502}%
  \BibitemOpen
  \bibfield  {author} {\bibinfo {author} {\bibfnamefont {A.~R.}\ \bibnamefont
  {Botello-M\'endez}}, \bibinfo {author} {\bibfnamefont {J.~C.}\ \bibnamefont
  {Obeso-Jureidini}}, \ and\ \bibinfo {author} {\bibfnamefont {G.~G.}\
  \bibnamefont {Naumis}},\ }\href {\doibase 10.1021/acs.jpcc.8b04502}
  {\bibfield  {journal} {\bibinfo  {journal} {The Journal of Physical Chemistry
  C}\ }\textbf {\bibinfo {volume} {122}},\ \bibinfo {pages} {15753} (\bibinfo
  {year} {2018})}\BibitemShut {NoStop}%
\bibitem [{\citenamefont {Mohiuddin}\ \emph {et~al.}(2009)\citenamefont
  {Mohiuddin}, \citenamefont {Lombardo}, \citenamefont {Nair}, \citenamefont
  {Bonetti}, \citenamefont {Savini}, \citenamefont {Jalil}, \citenamefont
  {Bonini}, \citenamefont {Basko}, \citenamefont {Galiotis}, \citenamefont
  {Marzari}, \citenamefont {Novoselov}, \citenamefont {Geim},\ and\
  \citenamefont {Ferrari}}]{PhysRevB.79.205433}%
  \BibitemOpen
  \bibfield  {author} {\bibinfo {author} {\bibfnamefont {T.~M.~G.}\
  \bibnamefont {Mohiuddin}}, \bibinfo {author} {\bibfnamefont {A.}~\bibnamefont
  {Lombardo}}, \bibinfo {author} {\bibfnamefont {R.~R.}\ \bibnamefont {Nair}},
  \bibinfo {author} {\bibfnamefont {A.}~\bibnamefont {Bonetti}}, \bibinfo
  {author} {\bibfnamefont {G.}~\bibnamefont {Savini}}, \bibinfo {author}
  {\bibfnamefont {R.}~\bibnamefont {Jalil}}, \bibinfo {author} {\bibfnamefont
  {N.}~\bibnamefont {Bonini}}, \bibinfo {author} {\bibfnamefont {D.~M.}\
  \bibnamefont {Basko}}, \bibinfo {author} {\bibfnamefont {C.}~\bibnamefont
  {Galiotis}}, \bibinfo {author} {\bibfnamefont {N.}~\bibnamefont {Marzari}},
  \bibinfo {author} {\bibfnamefont {K.~S.}\ \bibnamefont {Novoselov}}, \bibinfo
  {author} {\bibfnamefont {A.~K.}\ \bibnamefont {Geim}}, \ and\ \bibinfo
  {author} {\bibfnamefont {A.~C.}\ \bibnamefont {Ferrari}},\ }\href {\doibase
  10.1103/PhysRevB.79.205433} {\bibfield  {journal} {\bibinfo  {journal} {Phys.
  Rev. B}\ }\textbf {\bibinfo {volume} {79}},\ \bibinfo {pages} {205433}
  (\bibinfo {year} {2009})}\BibitemShut {NoStop}%
\bibitem [{\citenamefont {Ding}\ \emph {et~al.}(2010)\citenamefont {Ding},
  \citenamefont {Ji}, \citenamefont {Chen}, \citenamefont {Herklotz},
  \citenamefont {Dörr}, \citenamefont {Mei}, \citenamefont {Rastelli},\ and\
  \citenamefont {Schmidt}}]{doi:10.1021/nl101533x}%
  \BibitemOpen
  \bibfield  {author} {\bibinfo {author} {\bibfnamefont {F.}~\bibnamefont
  {Ding}}, \bibinfo {author} {\bibfnamefont {H.}~\bibnamefont {Ji}}, \bibinfo
  {author} {\bibfnamefont {Y.}~\bibnamefont {Chen}}, \bibinfo {author}
  {\bibfnamefont {A.}~\bibnamefont {Herklotz}}, \bibinfo {author}
  {\bibfnamefont {K.}~\bibnamefont {Dörr}}, \bibinfo {author} {\bibfnamefont
  {Y.}~\bibnamefont {Mei}}, \bibinfo {author} {\bibfnamefont {A.}~\bibnamefont
  {Rastelli}}, \ and\ \bibinfo {author} {\bibfnamefont {O.~G.}\ \bibnamefont
  {Schmidt}},\ }\href {\doibase 10.1021/nl101533x} {\bibfield  {journal}
  {\bibinfo  {journal} {Nano Letters}\ }\textbf {\bibinfo {volume} {10}},\
  \bibinfo {pages} {3453} (\bibinfo {year} {2010})}\BibitemShut {NoStop}%
\bibitem [{\citenamefont {Verberck}\ \emph {et~al.}(2012)\citenamefont
  {Verberck}, \citenamefont {Partoens}, \citenamefont {Peeters},\ and\
  \citenamefont {Trauzettel}}]{PhysRevB.85.125403}%
  \BibitemOpen
  \bibfield  {author} {\bibinfo {author} {\bibfnamefont {B.}~\bibnamefont
  {Verberck}}, \bibinfo {author} {\bibfnamefont {B.}~\bibnamefont {Partoens}},
  \bibinfo {author} {\bibfnamefont {F.~M.}\ \bibnamefont {Peeters}}, \ and\
  \bibinfo {author} {\bibfnamefont {B.}~\bibnamefont {Trauzettel}},\ }\href
  {\doibase 10.1103/PhysRevB.85.125403} {\bibfield  {journal} {\bibinfo
  {journal} {Phys. Rev. B}\ }\textbf {\bibinfo {volume} {85}},\ \bibinfo
  {pages} {125403} (\bibinfo {year} {2012})}\BibitemShut {NoStop}%
\bibitem [{\citenamefont {Mikitik}\ and\ \citenamefont
  {Sharlai}(2008)}]{PhysRevB.77.113407}%
  \BibitemOpen
  \bibfield  {author} {\bibinfo {author} {\bibfnamefont {G.~P.}\ \bibnamefont
  {Mikitik}}\ and\ \bibinfo {author} {\bibfnamefont {Y.~V.}\ \bibnamefont
  {Sharlai}},\ }\href {\doibase 10.1103/PhysRevB.77.113407} {\bibfield
  {journal} {\bibinfo  {journal} {Phys. Rev. B}\ }\textbf {\bibinfo {volume}
  {77}},\ \bibinfo {pages} {113407} (\bibinfo {year} {2008})}\BibitemShut
  {NoStop}%
\bibitem [{\citenamefont {Battilomo}\ \emph {et~al.}(2019)\citenamefont
  {Battilomo}, \citenamefont {Scopigno},\ and\ \citenamefont
  {Ortix}}]{PhysRevLett.123.196403}%
  \BibitemOpen
  \bibfield  {author} {\bibinfo {author} {\bibfnamefont {R.}~\bibnamefont
  {Battilomo}}, \bibinfo {author} {\bibfnamefont {N.}~\bibnamefont {Scopigno}},
  \ and\ \bibinfo {author} {\bibfnamefont {C.}~\bibnamefont {Ortix}},\ }\href
  {\doibase 10.1103/PhysRevLett.123.196403} {\bibfield  {journal} {\bibinfo
  {journal} {Phys. Rev. Lett.}\ }\textbf {\bibinfo {volume} {123}},\ \bibinfo
  {pages} {196403} (\bibinfo {year} {2019})}\BibitemShut {NoStop}%
\bibitem [{\citenamefont {Varlet}\ \emph {et~al.}(2014)\citenamefont {Varlet},
  \citenamefont {Bischoff}, \citenamefont {Simonet}, \citenamefont {Watanabe},
  \citenamefont {Taniguchi}, \citenamefont {Ihn}, \citenamefont {Ensslin},
  \citenamefont {Mucha-Kruczy\ifmmode~\acute{n}\else \'{n}\fi{}ski},\ and\
  \citenamefont {Fal'ko}}]{PhysRevLett.113.116602}%
  \BibitemOpen
  \bibfield  {author} {\bibinfo {author} {\bibfnamefont {A.}~\bibnamefont
  {Varlet}}, \bibinfo {author} {\bibfnamefont {D.}~\bibnamefont {Bischoff}},
  \bibinfo {author} {\bibfnamefont {P.}~\bibnamefont {Simonet}}, \bibinfo
  {author} {\bibfnamefont {K.}~\bibnamefont {Watanabe}}, \bibinfo {author}
  {\bibfnamefont {T.}~\bibnamefont {Taniguchi}}, \bibinfo {author}
  {\bibfnamefont {T.}~\bibnamefont {Ihn}}, \bibinfo {author} {\bibfnamefont
  {K.}~\bibnamefont {Ensslin}}, \bibinfo {author} {\bibfnamefont
  {M.}~\bibnamefont {Mucha-Kruczy\ifmmode~\acute{n}\else \'{n}\fi{}ski}}, \
  and\ \bibinfo {author} {\bibfnamefont {V.~I.}\ \bibnamefont {Fal'ko}},\
  }\href {\doibase 10.1103/PhysRevLett.113.116602} {\bibfield  {journal}
  {\bibinfo  {journal} {Phys. Rev. Lett.}\ }\textbf {\bibinfo {volume} {113}},\
  \bibinfo {pages} {116602} (\bibinfo {year} {2014})}\BibitemShut {NoStop}%
\bibitem [{\citenamefont {Novoselov}\ \emph {et~al.}(2006)\citenamefont
  {Novoselov}, \citenamefont {McCann}, \citenamefont {Morozov}, \citenamefont
  {Fal'ko}, \citenamefont {Katsnelson}, \citenamefont {Zeitler}, \citenamefont
  {Jiang}, \citenamefont {Schedin},\ and\ \citenamefont
  {Geim}}]{Novoselov2006}%
  \BibitemOpen
  \bibfield  {author} {\bibinfo {author} {\bibfnamefont {K.~S.}\ \bibnamefont
  {Novoselov}}, \bibinfo {author} {\bibfnamefont {E.}~\bibnamefont {McCann}},
  \bibinfo {author} {\bibfnamefont {S.~V.}\ \bibnamefont {Morozov}}, \bibinfo
  {author} {\bibfnamefont {V.~I.}\ \bibnamefont {Fal'ko}}, \bibinfo {author}
  {\bibfnamefont {M.~I.}\ \bibnamefont {Katsnelson}}, \bibinfo {author}
  {\bibfnamefont {U.}~\bibnamefont {Zeitler}}, \bibinfo {author} {\bibfnamefont
  {D.}~\bibnamefont {Jiang}}, \bibinfo {author} {\bibfnamefont
  {F.}~\bibnamefont {Schedin}}, \ and\ \bibinfo {author} {\bibfnamefont
  {A.~K.}\ \bibnamefont {Geim}},\ }\href {\doibase 10.1038/nphys245} {\bibfield
   {journal} {\bibinfo  {journal} {Nature Physics}\ }\textbf {\bibinfo {volume}
  {2}},\ \bibinfo {pages} {177} (\bibinfo {year} {2006})}\BibitemShut {NoStop}%
\bibitem [{2ba()}]{2band}%
  \BibitemOpen
  \href@noop {} {}\bibinfo {note} {The two-band model for strained bilayer is
  given by ${H = \begin{pmatrix} - \frac{1}{2} \Delta & -(\pi^\dag)^2 / 2m +
  w_3 \\ -\pi^2 / 2m + w_3^* & \frac{1}{2} \Delta \end{pmatrix}}$, where ${m =
  \gamma_1 / 2 v_0^2}$ and ${w_3 = |w_3| e^{i \phi}}$. In this case, we find
  band edges at $\bm{p} = \pm \sqrt{2 m |w_3|} (\xi \cos(\frac{\phi}{2}),
  -\sin(\frac{\phi}{2}))$ and $\epsilon = \pm \frac{1}{2} \Delta$. At the band
  edge, the magnetic moment is given by ${\mu^{z,*} = \xi 2 e \hbar
  \frac{|w_3|}{m \Delta}}$, leading to the estimation ${g_v^* \sim - 100 \xi
  \frac{|w_3|}{\Delta}}$.}\BibitemShut {Stop}%
\bibitem [{LL()}]{LL}%
  \BibitemOpen
  \href@noop {} {}\bibinfo {note} {Here we use the basis of LL orbitals
  $\ket{n} \, (n=0,1, \cdots)$, such that ${\pi \ket{n} = -i \sqrt{2 n \hbar e
  B} \ket{n - 1}}$ and ${\pi^\dag \ket{n} = i \sqrt{2 (n + 1) \hbar e B} \ket{n
  + 1}}$ in the $\bm{K}^+$ valley~\cite{McCann2013}. In the $\bm{K}^-$ valley,
  the roles are reversed: ${\pi \ket{n} = -i \sqrt{2 (n + 1) \hbar e B} \ket{n
  + 1}}$ and ${\pi^\dag \ket{n} = i \sqrt{2 n \hbar |e| B} \ket{n - 1}}$. We
  then expand Eq.~(\ref{eq:H}) into a ${4N \times 4N}$ basis, where each of the
  sublattices $(A,B',A',B)$ has a total of ${(N + 1, N, N, N - 1)}$ LLs
  considered in the $\bm{K}^+$ valley (${(N - 1, N, N, N + 1)}$ in the
  $\bm{K}^-$ valley). In Fig.~\ref{fig:LL} and Fig.~\ref{fig:LLK}, we used $N =
  500$.}\BibitemShut {Stop}%
\bibitem [{\citenamefont {Horsell}\ \emph {et~al.}(2008)\citenamefont
  {Horsell}, \citenamefont {Tikhonenko}, \citenamefont {Gorbachev},\ and\
  \citenamefont {Savchenko}}]{doi:10.1098/rsta.2007.2159}%
  \BibitemOpen
  \bibfield  {author} {\bibinfo {author} {\bibfnamefont {D.}~\bibnamefont
  {Horsell}}, \bibinfo {author} {\bibfnamefont {F.}~\bibnamefont {Tikhonenko}},
  \bibinfo {author} {\bibfnamefont {R.}~\bibnamefont {Gorbachev}}, \ and\
  \bibinfo {author} {\bibfnamefont {A.}~\bibnamefont {Savchenko}},\ }\href
  {\doibase 10.1098/rsta.2007.2159} {\bibfield  {journal} {\bibinfo  {journal}
  {Philosophical Transactions of the Royal Society A: Mathematical, Physical
  and Engineering Sciences}\ }\textbf {\bibinfo {volume} {366}},\ \bibinfo
  {pages} {245} (\bibinfo {year} {2008})}\BibitemShut {NoStop}%
\bibitem [{Kop()}]{Koppens}%
  \BibitemOpen
  \href@noop {} {}\bibinfo {note} {Private communication with F.
  Koppens}\BibitemShut {NoStop}%
\bibitem [{\citenamefont {Fuchs}\ \emph {et~al.}(2010)\citenamefont {Fuchs},
  \citenamefont {Pi{\'e}chon}, \citenamefont {Goerbig},\ and\ \citenamefont
  {Montambaux}}]{Fuchs2010}%
  \BibitemOpen
  \bibfield  {author} {\bibinfo {author} {\bibfnamefont {J.~N.}\ \bibnamefont
  {Fuchs}}, \bibinfo {author} {\bibfnamefont {F.}~\bibnamefont {Pi{\'e}chon}},
  \bibinfo {author} {\bibfnamefont {M.~O.}\ \bibnamefont {Goerbig}}, \ and\
  \bibinfo {author} {\bibfnamefont {G.}~\bibnamefont {Montambaux}},\ }\href
  {\doibase 10.1140/epjb/e2010-00259-2} {\bibfield  {journal} {\bibinfo
  {journal} {The European Physical Journal B}\ }\textbf {\bibinfo {volume}
  {77}},\ \bibinfo {pages} {351} (\bibinfo {year} {2010})}\BibitemShut
  {NoStop}%
\bibitem [{\citenamefont {Kuzmenko}\ \emph {et~al.}(2009)\citenamefont
  {Kuzmenko}, \citenamefont {Crassee}, \citenamefont {van~der Marel},
  \citenamefont {Blake},\ and\ \citenamefont {Novoselov}}]{PhysRevB.80.165406}%
  \BibitemOpen
  \bibfield  {author} {\bibinfo {author} {\bibfnamefont {A.~B.}\ \bibnamefont
  {Kuzmenko}}, \bibinfo {author} {\bibfnamefont {I.}~\bibnamefont {Crassee}},
  \bibinfo {author} {\bibfnamefont {D.}~\bibnamefont {van~der Marel}}, \bibinfo
  {author} {\bibfnamefont {P.}~\bibnamefont {Blake}}, \ and\ \bibinfo {author}
  {\bibfnamefont {K.~S.}\ \bibnamefont {Novoselov}},\ }\href {\doibase
  10.1103/PhysRevB.80.165406} {\bibfield  {journal} {\bibinfo  {journal} {Phys.
  Rev. B}\ }\textbf {\bibinfo {volume} {80}},\ \bibinfo {pages} {165406}
  (\bibinfo {year} {2009})}\BibitemShut {NoStop}%
\bibitem [{\citenamefont {McCann}\ and\ \citenamefont
  {Koshino}(2013)}]{McCann2013}%
  \BibitemOpen
  \bibfield  {author} {\bibinfo {author} {\bibfnamefont {E.}~\bibnamefont
  {McCann}}\ and\ \bibinfo {author} {\bibfnamefont {M.}~\bibnamefont
  {Koshino}},\ }\href {\doibase 10.1088/0034-4885/76/5/056503} {\bibfield
  {journal} {\bibinfo  {journal} {Reports on Progress in Physics}\ }\textbf
  {\bibinfo {volume} {76}},\ \bibinfo {pages} {056503} (\bibinfo {year}
  {2013})}\BibitemShut {NoStop}%
\end{thebibliography}%

\clearpage
\appendix

\end{document}